\documentclass{aastex63}

\newcommand{\fpp}[2]{\frac{\partial #1}{\partial #2}}
\newcommand{\vctr}[1]{\mbox{\boldmath $#1$}}

\received{November, 12, 2020}
\revised{January 10, 2021}
\accepted{\today}
\submitjournal{ApJ}

\shorttitle{data-driven}
\shortauthors{Kaneko et al.}

\begin{document}

\title{Data-driven MHD simulation of successive solar plasma eruptions}

\author[0000-0002-7800-9262]{Takafumi Kaneko}
\affiliation{Institute for Space-Earth Environmental Research, Nagoya University,\\
Furo-cho, Chikusa-ku, Nagoya, Aichi, 464-8601, Japan}

\author[0000-0001-9149-6547]{Sung-Hong Park}
\affiliation{Institute for Space-Earth Environmental Research, Nagoya University,\\
Furo-cho, Chikusa-ku, Nagoya, Aichi, 464-8601, Japan}

\author[0000-0002-6814-6810]{Kanya Kusano}
\affiliation{Institute for Space-Earth Environmental Research, Nagoya University,\\
Furo-cho, Chikusa-ku, Nagoya, Aichi, 464-8601, Japan}

\correspondingauthor{Takafumi Kaneko}
\email{kaneko@isee.nagoya-u.ac.jp}



\begin{abstract}

Solar flares and plasma eruptions are sudden releases of magnetic energy stored in the plasma atmosphere. To understand the physical mechanisms governing their occurrences, three-dimensional magnetic fields from
the photosphere up to the corona must be studied. The solar photospheric magnetic fields are 
observable, whereas the coronal magnetic fields cannot be measured. One method for inferring coronal magnetic fields is performing data-driven simulations,
which involves time-series observational data of the photospheric magnetic fields
with the bottom boundary of magnetohydrodynamic simulations.
We developed a data-driven method in which temporal evolutions of the observational vector magnetic
field can be reproduced at the bottom boundary in the simulation by introducing an inverted velocity field.
This velocity field is obtained by inversely solving the induction equation and applying an appropriate gauge transformation.
Using this method, we performed a data-driven simulation of successive small eruptions
observed by the Solar Dynamics Observatory and the Solar Magnetic Activity Telescope in November 2017.
The simulation well reproduced the converging motion between opposite-polarity magnetic patches,
demonstrating successive formation and eruptions of helical flux ropes.

\end{abstract}



\section{Introduction}\label{sec:intro}
Solar flares are the sudden releases of energy from the sun.
Flares are often accompanied by plasma eruptions such as prominence eruptions and coronal mass ejections.
Flares and plasma eruptions are caused by the release of magnetic energy
stored in the plasma atmosphere.
 Evidences of flares and plasma eruptions have also been found in other sun-like stars \citep[][]{Osten2005ApJ,PandeySingh2008MNRAS,Maehara2012Natur,Notsu2019ApJ,Namekata2020PASJ}.
The sun is the only star of which the photospheric magnetic 
fields can be observed with a high spatio-temporal resolution. From the solar observations and theoretical studies based on magnetohydrodynamic (MHD) theories, we can infer the detailed magnetic activities leading to the sudden energy release, 
which can be common in other sun-like stars.
In the current understanding of solar physics,
magnetic reconnection and MHD instabilities
\citep[][]{HoodPriest1979SoPh,KliemTorok2006PhRvL,IshiguroKusano2017ApJ}
are the essential mechanisms leading to the magnetic energy release. 

In the solar observations, the photospheric magnetic fields are temporally
changed via advection by convective flows
and magnetic fluxes emerging from the deeper convection zone.  
To reveal the mechanisms of the explosive events
and develop methodologies to predict them,
previous studies attempted to evaluate
the possibility of magnetic reconnection and
the critical conditions of MHD instabilities \citep[][]{Amari2014Natur,Kusano2020Sci}.
For this, the information of three-dimensional magnetic fields
from the photosphere up to the corona is required. 
The photospheric magnetic fields can be observed,
whereas the coronal magnetic fields cannot be measured directly.
Previous studies developed numerical methods to extrapolate
three-dimensional coronal magnetic fields from the
two-dimensional observational vector magnetic fields in the photosphere,
e.g., nonlinear force-free field (NLFFF)
 approximation \citep[reviewed by][]{Inoue2016PEPS}.
There have been attempts of data-constrained simulations, where the NLFFF approximation
was used as the initial condition of MHD simulation
\citep[][]{Amari2014Natur,Muhamad2017ApJ}.
In these simulations, the photospheric magnetic fields after
temporal integration did not always reproduce the observed ones.
Another attempt was the data-driven simulation in which
time-series photospheric magnetic data were involved in the
bottom boundary of MHD simulations.
The expected advantage of the data-driven methods, compared with the NLFFF
or data-constrained model, is that the results are free from the assumption of force-free field.
We can follow more realistic temporal evolution of coronal magnetic
fields as a response of temporal change of the observational photospheric
magnetic fields. Several data-driven MHD simulations have been
performed, and their results agree with some aspects
in the observations, e.g., morphology of the coronal magnetic loops
\citep{CheungDerosa2012ApJ,CheungDerosa2015ApJ,Jiang2016ApJ,Hayashi2018ApJ,Hayashi2019ApJL,Pomoell2019SoPh,GuoYang2019ApJ,He2020ApJ}.
In contrast, a recent comparative study
by \citet{Toriumi2020ApJ} reported that the numerical solutions
obtained from the different data-driven simulations using the same
time-series magnetic data were different from each other.
The data-driven methods must be improved further to resolve these discrepancies.

In this study, we focus on the velocity fields in the bottom boundary
of MHD simulation. In several data-driven methods, the velocity
fields at the bottom boundary were set to be zero, leading to
physical inconsistency between velocity fields and electric
or magnetic fields in terms of the induction equation.
A recent study by \citet{Hayashi2019ApJL} combined 
the velocity fields derived from differential affine velocity estimator for vector magnetograms
\citep[DAVE4VM;][]{Schuck2006ApJ}
with their own data-driven method (denoted as the $v$-driven method in their paper).
They confirmed that the frozen-in condition between plasmas and magnetic fields was well
established. This is because the DAVE4VM-inferred velocity works as the bottom boundary
condition to the equation of motion in MHD simulation, providing the motion of plasmas
coherent with the time evolution of the magnetic fields in the observation.
Another recent study by \citet{GuoYang2019ApJ} reported that the numerical
results of data-driven simulations with and without velocity fields by DAVE4VM were 
similar in terms of morphology and propagation path of the erupted flux ropes.
They argued that eruption inevitably happens if the initial condition of MHD simulation is already close to the dynamic eruptive phase.  Their conclusion was that the change of the bottom boundary condition had subtle effect to the onset mechanism, while it would affect magnetic energy build-up before eruptive phase. Since the observational targets and many aspects of numerical techniques were different in \citet{Hayashi2019ApJL} and \citet{GuoYang2019ApJ}, it is fairly difficult to compare their results. 
One issue we concern about their numerical techniques is that the DAVE4VM-inferred velocity is not always consistent with the inverted electric fields or the time evolution of the observational magnetic fields used as the bottom boundary condition of MHD simulations.
In the present study, we attempted to implement an inversion technique
of the induction equation directly in our simulation code,
and proposed a method to derive the velocity fields
reproducing the observed time evolution of magnetic field as a numerical solution of MHD equations.
To confirm the feasibility of the method, we applied it
to the successive small eruptive events that occurred in November 2017.

The observation of the eruptive events are described in Section \ref{sec:obs}.
The numerical method including velocity inversion is described in Section \ref{sec:sim}.
The numerical results are shown in Section \ref{sec:res}.
We summarize and discuss the results in Section \ref{sec:sum} 

\section{Observation} \label{sec:obs}
\begin{figure}[htbp]
  \begin{center}
    \includegraphics[width=1\textwidth]{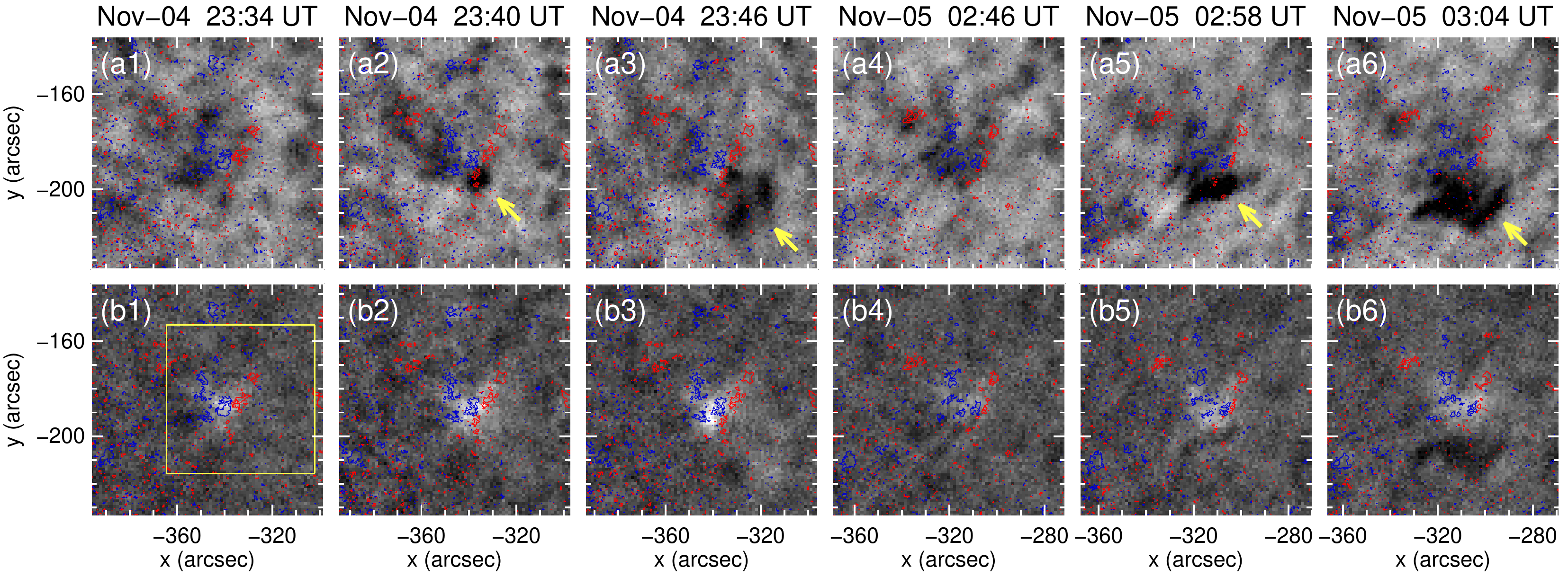}
    \caption{Two successive eruption events observed in H$\mathrm{\alpha}-$0.5\,{\AA} images (top panels) and H$\mathrm{\alpha}$ line center images (bottom panels). The arrows in panels (a2) and (a3) indicate the first event, while those in panels (a5) and (a6) indicate the second event. In each panel, the red and blue contours represent $\pm$50 G of the vertical magnetic field $B_{z}$. The yellow box in panel (b1) represents the region used as the bottom boundary in our MHD simulation.}
    \label{fig:ha}
  \end{center}
\end{figure}

The Solar Dynamics Doppler Imager \citep[SDDI;][]{Ichimoto2017_SDDI_SoPh} installed on the Solar Magnetic Activity Research Telescope \citep[SMART;][]{Ueno2004_SMART_SPIE} at Hida Observatory of Kyoto University provides full-disk solar images at multiple wavelengths around the H$\mathrm{\alpha}$ 6563\,{\AA} line with a 0.25\,{\AA} bandpass.
The top and bottom panels of Figure \ref{fig:ha} show H$\mathrm{\alpha}$ blue wing images at $-$0.5\,{\AA} from the line center and H$\mathrm{\alpha}$ line center images, respectively, in six snapshots taken from SMART/SDDI observations on November 4--5, 2017, which demonstrate two successive eruptions. As indicated by the arrow in panel (a2) of Figure \ref{fig:ha}, the first eruption event started at 23:40 UT on November 4 and appeared as a compact dark feature with a size of $\sim$10{\arcsec} in H$\mathrm{\alpha}-$0.5\,{\AA}. This dark feature (i.e., a so-called H$\mathrm{\alpha}$ upflow event), which is only visible in the H$\mathrm{\alpha}$ blue wing, displays an upward motion and is known to be often associated with magnetic reconnection \citep[][]{1998SoPh..178...55W,1998ApJ...504L.123C}. In a sequence of H$\mathrm{\alpha}-$0.5\,{\AA} images, it was found that the H$\mathrm{\alpha}-$0.5\,{\AA} upflow features increase in size as they erupt in the south-west direction (see panel (a3)). During the eruption, an enhanced brightening in H$\mathrm{\alpha}$ was observed near the magnetic polarity inversion line (PIL), where two opposite-polarity magnetic patches approached each other. Approximately 3 h after the first eruption, another eruption event began at 02:50 UT on November 5, exhibiting characteristics similar to those of the first eruption event in the context of the south-west eruption direction and H$\mathrm{\alpha}$ brightening. In the case of the second eruption, contrarily, we note that an inverse S-shaped structure is clearly seen in the H$\mathrm{\alpha}$ line center (refer to panel (b6)).

\begin{figure}[htbp]
  \begin{center}
    \includegraphics[scale=1.7]{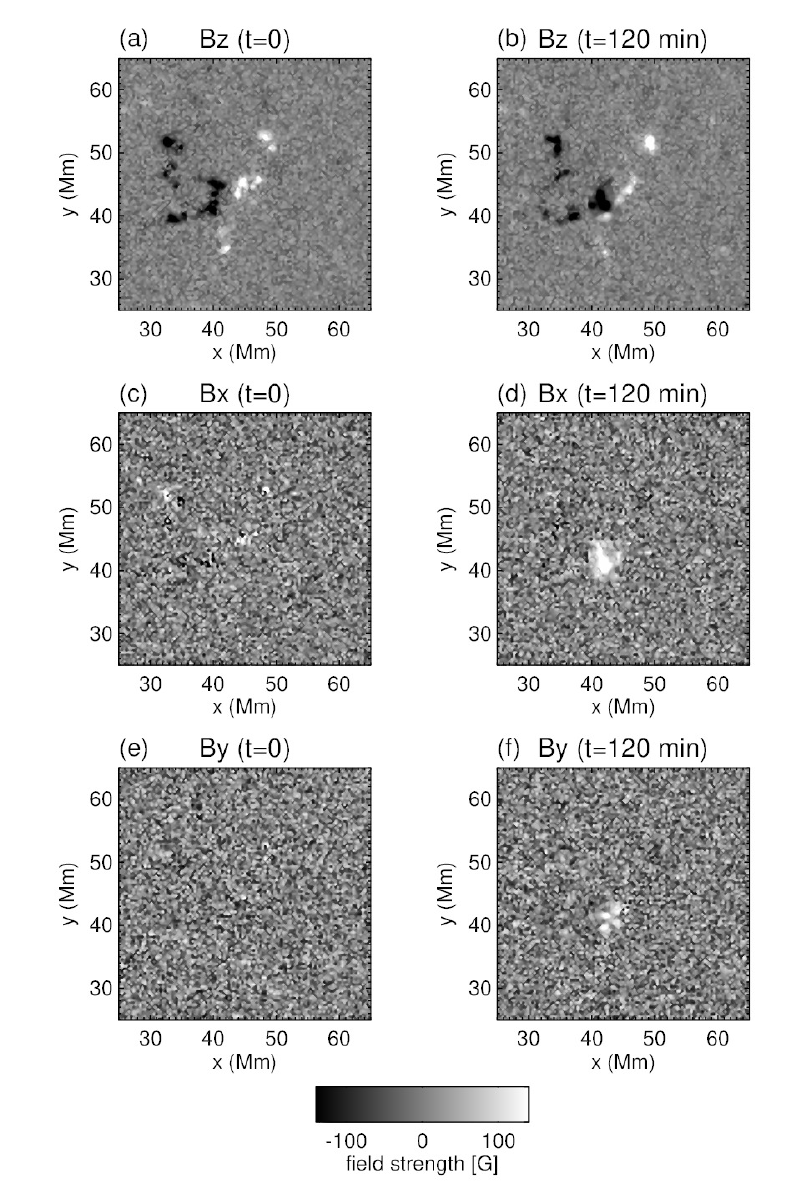}
    \caption{Snapshots of time-series data of magnetic field observed by SDO/HMI.
      Panels (a), (c), and (e) show snapshots of 22:58 UT on November 4, 2017 (corresponding to $t=0$ in the simulation).
      Panels (b), (d), and (f) show snapshots of 00:58 UT on November 5, 2017 (corresponding to $t=120~\mathrm{min}$
      in the simulation). The field-of-view of these figures is represented by the yellow box in
      Fig. \ref{fig:ha} (b1)}
    \label{fig:hmi}
  \end{center}
\end{figure}

In this study, we used a sequence of photospheric vector magnetograms obtained at 12-min cadence by the Helioseismic and Magnetic Imager \citep[HMI;][]{Schou2012_HMI_SoPh} onboard the Solar Dynamics Observatory \citep[SDO;][]{Pesnell2012_SDO_SoPh}. The pixel size of the HMI vector magnetograms was $\sim$360\,km. Figure \ref{fig:hmi} shows two co-aligned images of the vertical ($B_{z}$) and horizontal ($B_{x}$ and $B_{y}$) components of the photospheric magnetic field at 22:58 UT on November 4, 2017 (left column) and at 00:58 UT on November 5, 2017 (right column). The field-of-view of the co-aligned magnetic field images is marked by the yellow box in panel (b1) of Figure \ref{fig:ha}, which contains the magnetic source region that produced the two eruptions. The source region consists of two main opposite-polarity magnetic patches that, in general, showed a converging motion as well as a decrease in the magnetic flux of both polarities over a 5 h interval around the times that the two eruptions occurred. Moreover, as shown in panels (d) and (f) of Figure \ref{fig:hmi}, the strengths of the horizontal components $B_{x}$ and $B_{y}$ are found to increase after the first eruption.

\section{Numerical Method} \label{sec:sim}

We numerically solved the zero-beta MHD equations as follows.
\begin{equation}
  \fpp{\rho }{t}+\nabla \cdot \left(\rho \vctr{v} \right)=0,
  \label{eq:mass}
\end{equation}
\begin{equation}
  \fpp{\left(\rho \vctr{v} \right)}{t}+\nabla \cdot \left(\rho \vctr{v}\vctr{v} + \frac{B^{2}}{\
8\pi }\vctr{I}
  -\frac{\vctr{B}\vctr{B}}{4\pi } \right)=0,
\end{equation}
\begin{equation}
  \fpp{\vctr{B}}{t}=\nabla \times (\vctr{v}\times \vctr{B} -\eta \vctr{J}),
  \label{eq:induction}
\end{equation}
\begin{equation}
  \vctr{J}=\frac{1}{4\pi }\nabla \times \vctr{B},
  \label{eq:cur}
\end{equation}
where $t$, $\rho $, $\vctr{v}$, $\vctr{B}$, $\vctr{J}$, $\eta $, and $\vctr{I}$ denote
time, mass density, velocity fields, magnetic fields, current density, resistivity, and unit vector, respectively.
We used the anomalous resistivity in the following form:
\begin{equation}
  \eta=0,~~(J < J_{c}),
\end{equation}
\begin{equation}
  \eta=\eta _{0}(J/J_{c}-1)^{2},~~(J \geq J_{c}),
\end{equation}
where $J_{c}=10^{-9}~\mathrm{G/cm}$, $\eta _{0}=10^{11}~\mathrm{cm^{2}/s}$,
and we restrict $\eta \leq \eta _{\mathrm{max}}=10^{11}~\mathrm{cm^{2}/s}$.

We inverted velocity fields which reproduce the observational photospheric magnetic fields
by solving Eq. (\ref{eq:induction}), {and implemented them in the bottom boundary layer of the MHD simulation.}
The inverted velocity fields were computed by the following three steps:
\begin{enumerate}
\item Inversion of the induction equation\\
  We solved an inverse problem of the induction equation as, in principle, 
  \begin{equation}
    \frac{\vctr{B}_{\mathrm{obs}}^{n+1}-\vctr{B}_{\mathrm{obs}}^{n}}{\tau }=-\nabla \times \vctr{E}^{I},
    \label{eq:inv_one}
  \end{equation}
  where $\vctr{B}_{\mathrm{obs}}^{n}$, $\tau $, and $\vctr{E}^{I}$ represent the $n$-th snapshot in the
  time-series data of the observational magnetic fields, the temporal cadence of the HMI observation,
  and an inverted electric field, respectively.
  As pointed out in a previous study \citep[][]{Kusano2002ApJ}, we cannot solve this inverse problem completely
  because Eq. (\ref{eq:inv_one}) includes the derivative in the $z$-direction (the direction normal to the photosphere),
  whereas the observational magnetic data have only two-dimensional information in the $xy$-plane (corresponding to the
  solar surface).
  Several methods have been proposed to resolve this problem.
  In this study, we adopt the poloidal-toloidal decomposition method \citep{Fisher2010ApJ} and obtain
  $\vctr{E}^{I}$. The advantage of this method is that we can estimate the vertical derivative of
  electric fields to some extent. However, the complete solution cannot be obtained even by this method.
  We carried out the inversion of the electric fields
  between the simulated magnetic fields and the observational magnetic fields
  during the observational time cadence:
  \begin{equation}
    \frac{\vctr{B}_{\mathrm{obs}}^{n+1}-\vctr{B}_{\mathrm{sim}}^{n+m/M}}{(1-m/M)\tau }=-\nabla \times \vctr{E}^{I},
    \label{eq:inverse}
  \end{equation}
  where $\vctr{B}_{\mathrm{sim}}$ denotes the simulated magnetic fields, $m=1,2,...,M-1$ represents
  the $m$-th sub-snapshot between the $n$-th and the $(n+1)$-th observational snapshots.
  We adopted $M=6$ in this study, hence,
  the inversion was performed every 2 min during 12-min observational cadence of HMI.
  This piecewise inversion technique increases feasibility of the observational magnetic fields
  compared with the case that electric fields are inverted only once
  between $B_{\mathrm{obs}}^{n}$ and $B_{\mathrm{obs}}^{n+1}$.
\item Gauge transformation\\
  Electric field is mathematically gauge-invariant to the induction equation; 
  we can add an arbitrary scalar potential $\phi $ in the following form.
  \begin{equation}
    \vctr{E}=\vctr{E}^{I}-\nabla \phi,
  \end{equation}
  where $\vctr{E}$ represents the electric field after gauge transformation.
  In contrast, as demonstrated by \citet{Pomoell2019SoPh}, the results of data-driven simulations
  are influenced by gauge transformation. We adopted a gauge transformation that satisfied
  $\vctr{E} \cdot \vctr{B}=0$ using the iterative approach in \citet{Fisher2010ApJ}.
  The motivation to use this gauge transformation is as follows: the electric fields defined by
  $\vctr{E}=-\vctr{v}\times \vctr{B}$ are always perpendicular to magnetic fields.
  In contrast, the inverted electric fields $\vctr{E}^{I}$ before the gauge transformation
  usually contain nonzero $\vctr{E}_{\parallel }$ (parallel component to $\vctr{B}$).
  The nonzero $\vctr{E}_{\parallel }$ can cause mismatch of the electric fields
  between the bottom boundary layer and the main simulation domain
  because the electric fields in the main simulation domain are computed
  as $\vctr{E}=-\vctr{v}\times \vctr{B}$.
  Thus, we assumed that $\vctr{E} \cdot \vctr{B}=0$ is a necessary condition for the boundary
  electric fields in data-driven MHD simulations.
  Note that this assumption is valid even if resistive term $\eta \vctr{J}$ was introduced
  because the magnitude of $\eta \vctr{J}$ is constrained much smaller than that of
  $-\vctr{v}\times \vctr{B}$ in MHD simulations.
\item Derivation of velocity fields\\ 
  We compute velocity fields as follows:
  \begin{equation}
    \vctr{v}^{I}=\frac{\vctr{E} \times \vctr{B}}{B^{2}},
    \label{eq:exb}
  \end{equation}
  where $\vctr{v}^{I}$ represents the inverted velocity field.
  We substituted $\vctr{B}_{\mathrm{sim}}$ to $\vctr{B}$ in Eq. (\ref{eq:exb}).
  $\vctr{v}^{I}$ was updated every numerical time step.
\end{enumerate}
In Step 3, in the case that $\vctr{E}$ contains $\vctr{E}_{\parallel }$, $\vctr{v}^{I}$ loses the information of
$\vctr{E}_{\parallel }$ (because $\vctr{E}_{\parallel }\times \vctr{B}=0$).
In our manipulations, in Step 2, $\vctr{E}_{\parallel }$ has already been eliminated by the gauge transformation.
The inductive electric fields were calculated in a part of the right-hand side of Eq. (\ref{eq:induction}) as follows: 
\begin{equation}
  \vctr{v}^{I}\times \vctr{B} = \frac{(\vctr{E} \times \vctr{B})\times \vctr{B}}{B^{2}}
  =-\vctr{E}+\frac{(\vctr{E}\cdot \vctr{B})\vctr{B}}{B^{2}}=-\vctr{E}.
\end{equation}
Thus, we expect that the observed photospheric magnetic fields are
reproduced as a self-consistent numerical solution of Eq. (\ref{eq:induction})
only by introducing $\vctr{v}^{I}$ in the bottom boundary.
To reduce the observational noise in the area of weak magnetic fields,
which can damage the inversion of electric fields and the gauge transformation,
we applied a low-pass filter using FFT to the original magnetic data.
The practical spatial resolution was 8 times lower than the original spatial resolution of the HMI.

The simulation domain is a rectangular box. Its Cartesian coordinates $(x,y,z)$ are extended to
$0<x<89.6~\mathrm{Mm}$, $0<y<89.6~\mathrm{Mm}$, and $-1.44~\mathrm{Mm}<z<73.8~\mathrm{Mm}$, respectively,
where the $xy$-plane is the horizontal plane parallel to the solar surface,
and the $z$-direction represents the height.
We adopted uniform grid spacing in every direction, and the grid size was $360~\mathrm{km}$ corresponding to the spatial resolution of the HMI.
Below the $z=0$ plane, we set 5 grids in $z$-direction where 
Eq. (\ref{eq:induction}) was numerically solved by introducing $\vctr{v}^{I}$. Note that only the horizontal derivatives were calculated in the lowest 2 grids.
The $z=-360~\mathrm{km}$ plane (one grid below $z=0$) is at the height where
the observational magnetic fields were expected to be reproduced. The method of \citet{Fisher2010ApJ} derives $\partial _{z} E_{x}$ and $\partial _{z} E_{y}$. Assuming $\partial _{z} E_{z}=0$, we linearly extrapolated the inverted electric fields in the $z$-direction below $z=0$ and computed $\vctr{v}^{I}$ with the local magnetic fields using Eq. (\ref{eq:exb}). 
The density was fixed to the initial values below $z=0$.
We adopted free boundary condition to the top boundary and fixed to the side boundaries.
Our simulation only included the corona with typical density of 
$10^{9}~\mathrm{cm^{-3}}$, which is much smaller than the typical
photospheric density $10^{17}~\mathrm{cm^{-3}}$.
To suppress the unrealistically fast Alfv\`en speed,
we reduce the magnetic field strength to be 10 times smaller than
the original observed values. The same modification was also adopted in \citet{Jiang2016ApJ}.

The initial condition was a potential field computed by the Fourier expansion method \citep{Priest2014}
from the vertical magnetic field at 22:58 UT on November 4, 2017, observed by HMI.
The initial density was given by $\rho =\rho _{0}\exp [-z/H]$,
where $\rho _{0}=3.2 \times 10^{-15}~\mathrm{g/cm^{-3}}$ and $H=3.0\times 10^{4}~\mathrm{km}$. 
The numerical scheme used was a four-step Runge-Kutta method \citep{Jameson2017origins}
and a fourth order central finite difference method with an artificial viscosity \citep{Rempel2014ApJ}.

\section{Results} \label{sec:res}

\begin{figure}[htbp]
  \begin{center}
    \includegraphics[scale=1.7]{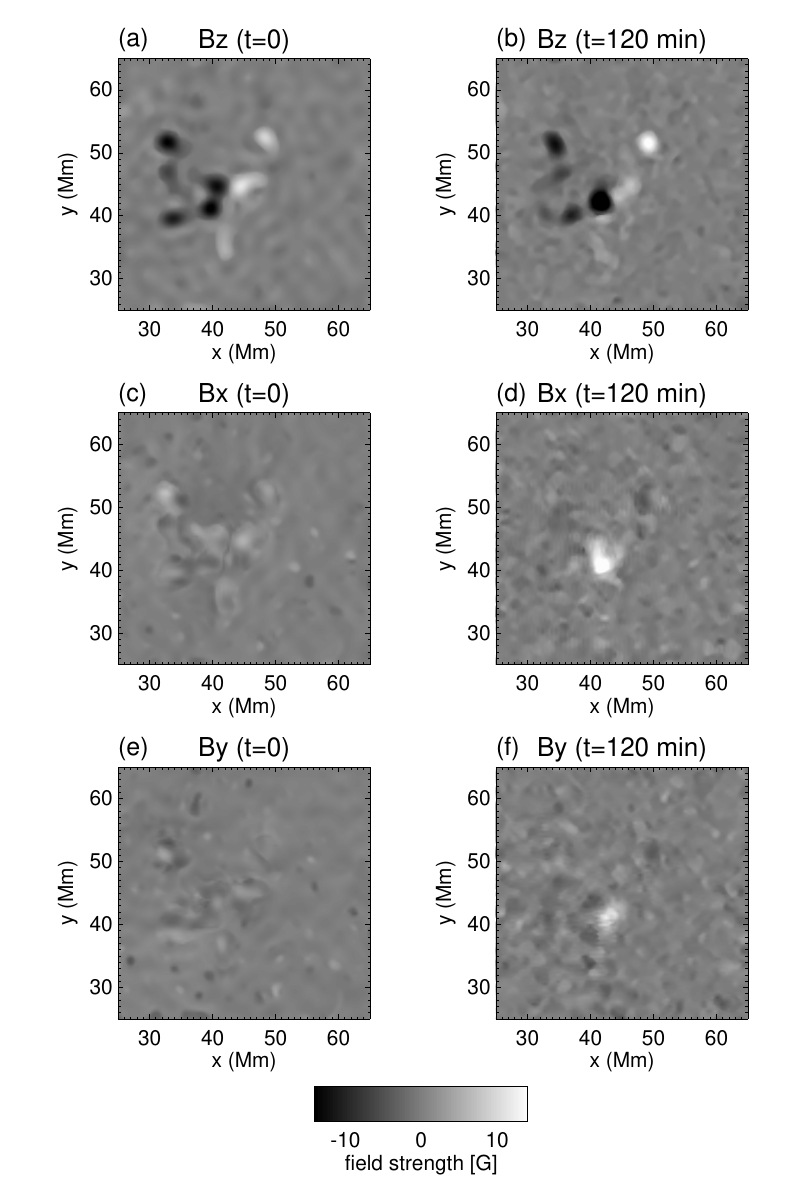}
    \caption{Snapshots of time evolution of magnetic fields at bottom boundary.
      Panels (a), (c), and (e) show snapshots of $t=0$ (corresponding to Nov. 4. 2017, 22:58 UT in the observation).
      Panels (b), (d), and (f) show snapshots of $t=120~\mathrm{min}$ (corresponding to Nov. 5. 2017, 0:58 UT
      in the observation).}
    \label{fig:sim_btm}
  \end{center}
\end{figure}

Figure \ref{fig:sim_btm} shows snapshots of magnetic fields in the bottom boundary
at the height $z=-360~\mathrm{km}$, where the observed magnetic fields were expected
to be reproduced.
Note that the magnetic fields shown in Fig. \ref{fig:sim_btm} are the numerically obtained solutions, not merely smoothed observational data.
Compared with Fig. \ref{fig:hmi},
we confirmed that the converging motion of the opposite-polarity magnetic patches and the intrusion of the negative patch to the positive patch were well reproduced in our simulation. The small structures were smoothed out by the low-pass filter used for the inversion and the anomalous resistivity during the temporal integration of the MHD simulation. 
The structural similarity (SSIM) values \citep[][]{Wang2004image} between the raw observational data and the low-pass filtered observational data 
at $t=120~\mathrm{min}$ were 0.22, 0.12, and 0.58 for the components $B_{x}$, $B_{y}$, and $B_{z}$, respectively, and the SSIM values 
between the low-pass filtered observational data and the simulated data were
0.82, 0.61, and 0.93 for the components $B_{x}$, $B_{y}$, and $B_{z}$, respectively. 
We used the low-pass filtered data for calculation of $\vctr{v}^{I}$.
We confirmed that the inverted velocities work well 
because the given magnetic fields were reproduced with high accuracy.

\begin{figure}[htbp]
  \begin{center}
    \includegraphics[scale=0.35]{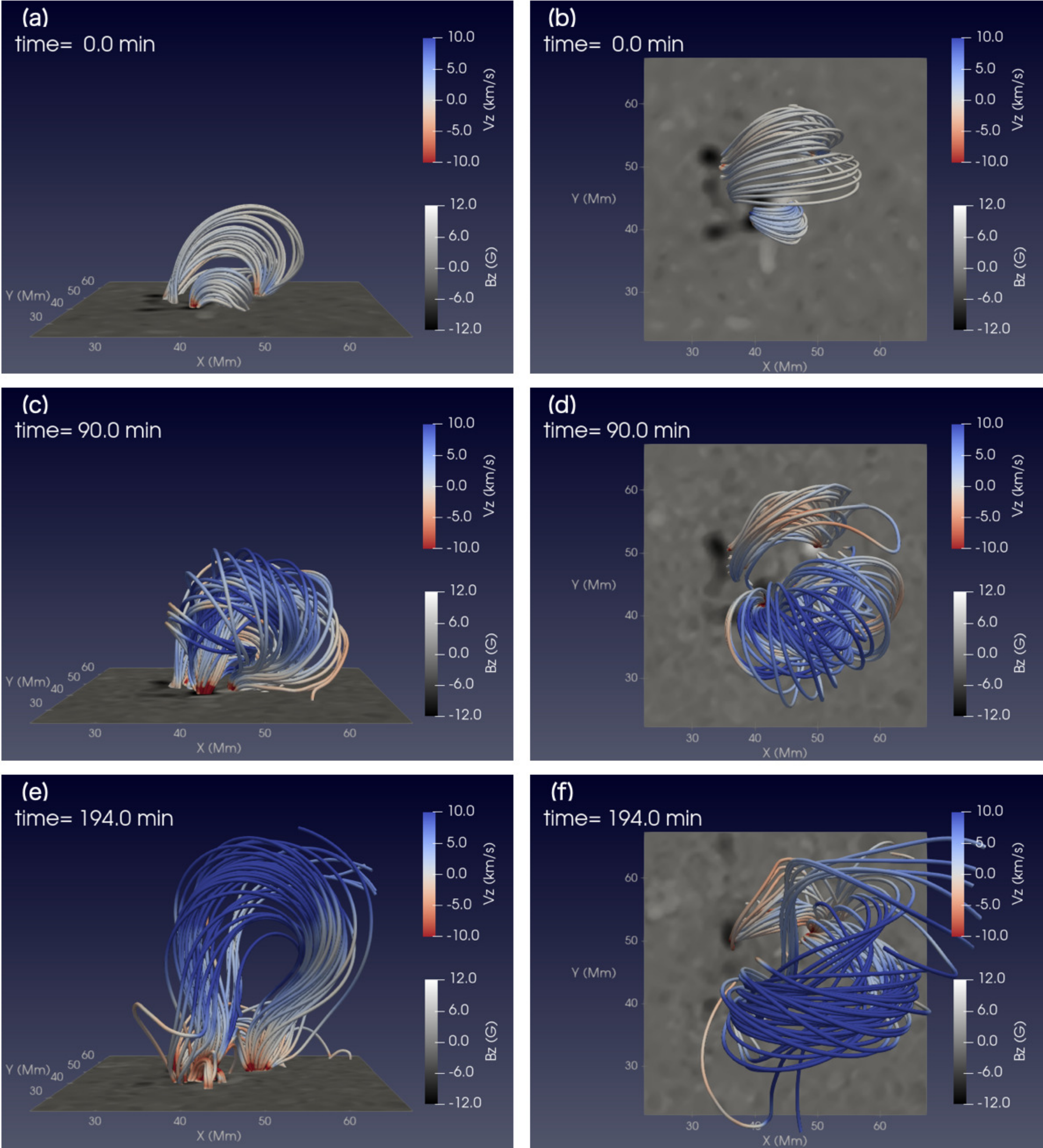}
    \caption{Temporal evolution of the coronal magnetic fields. Lines and colors on the lines
      represent the magnetic field lines and vertical velocity, respectively. Blue and red represent upward and downward velocities, 
  respectively. The grayscale on the bottom surface represents the vertical magnetic fields.}
    \label{fig:mag3d}
  \end{center}
\end{figure}

As the opposite-polarity patches converged with each other
and the negative magnetic patch further trespassed into the positive magnetic patch,
the formation and eruption of flux ropes via reconnection successively occurred in our simulation.
Figure \ref{fig:mag3d} shows the temporal evolution of the three-dimensional magnetic field in the corona.
Figures \ref{fig:mag3d} (a) and (b) show snapshots of when the horizontal magnetic fields were shifted
from the potential fields to the observed ones at 22:58 UT on November 4, 2017. Figures \ref{fig:mag3d} (c) and (d)
show snapshots of the first eruption. Figures \ref{fig:mag3d} (e) and (f)
show snapshots of the second eruption.
We succeeded in reproducing the successive eruptions of flux ropes.
In both cases, the flux ropes erupted in the south-west direction. 
We can interpret that the erupting filamentary structures in the observation were manifestations of the erupting flux ropes.
Compared with H$\mathrm{\alpha }$ blue wing images in the observation (see Fig. \ref{fig:ha} (a3) and (a6)),
the direction of eruptions in the simulation were in agreement with the observational results.

\begin{figure}[htbp]
  \begin{center}
    \includegraphics[scale=0.95]{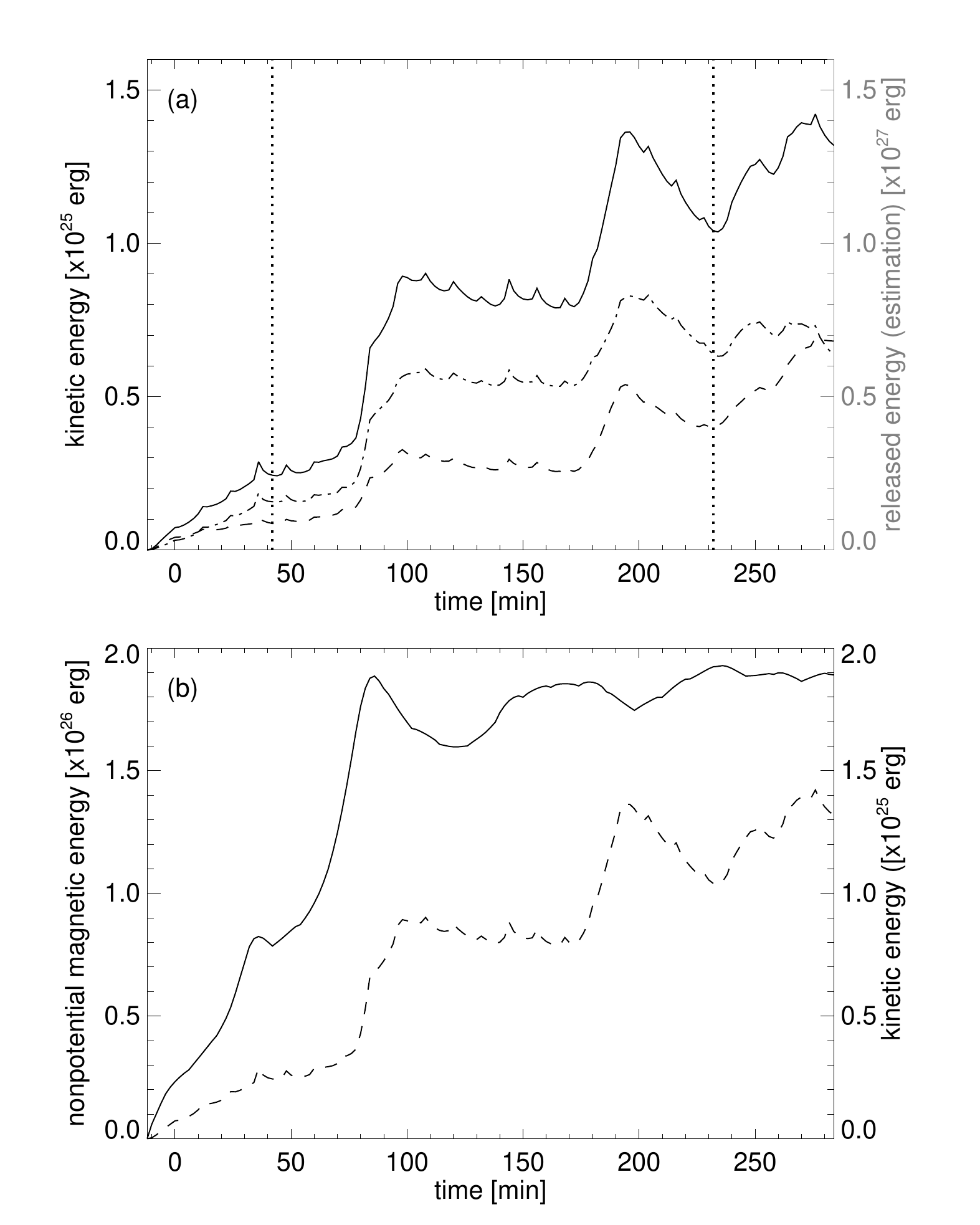}
    \caption{The solid, dashed, and dash-dotted lines in panel (a) represent
        the kinetic energy of all velocity components
        $\int _{z>0} \frac{1}{2}\rho (v_{x}^2+v_{y}^2+v_{z}^2)dV$,
        of the vertical component $\int _{z>0}\frac{1}{2}\rho v_{z}^2dV$,
        and of the horizontal components $\int _{z>0}\frac{1}{2}\rho (v_{x}^2+v_{y}^2)dV$, respectively.
        The vertical dotted lines indicate the actual onset times of the eruptions in the observation.
        The right axis in panel (a) represents the estimated released energy in reality.
        The solid and dashed lines in panel (b) represent the nonpotential magnetic energy
        and the kinetic energy of all velocity components, respectively.}
    \label{fig:kin}
  \end{center}
\end{figure}

Figure \ref{fig:kin} (a)
shows the temporal evolution of the kinetic energy integrated in the simulation
domain over the $z=0$ plane. The rapid increase in kinetic energy 
at $t\sim 100~\mathrm{min}$ and $t\sim 200~\mathrm{min} $ represents the eruptions. 
The onset time of the first eruption in the simulation was delayed $40~\mathrm{min}$ compared
with the observational results, and that of the second eruption was $40~\mathrm{min}$ earlier.
Figure \ref{fig:kin} (b) shows the temporal evolution of the nonpotential magnetic energy
computed as difference of the total magnetic energy and the potential magnetic energy.
The rapid increase of kinetic energy temporally coincided with
the reduction of nonpotential magnetic energy. 
The kinetic energy was approximately $1 \times 10^{25}~\mathrm{erg}$ and the released magnetic energy was $2-3 \times 10^{25}~\mathrm{erg}$. Note that the magnetic energy can change also due to the energy flux at the bottom and the top boundaries.
Because we reduced the magnetic field strength in the simulation to 10 times smaller than the original observational values, we can speculate that the actual energy release was of the order of $10^{27}~\mathrm{erg}$ for these eruptive events (the right axis of Fig. \ref{fig:kin} (a)). This is because magnetic energy is proportional to $B^{2}$, and we solved scale-free MHD equations.

\section{Summary and Discussion} \label{sec:sum}

We developed a numerical methodology that reproduces the temporal evolution of observational vector magnetic fields
in the bottom boundary of MHD simulation by introducing the velocity fields
inverted from the time-series observational magnetic data ($\vctr{E} \times \vctr{B}$-driven method).
The inverted velocity fields were computed by the formula of $\vctr{E} \times \vctr{B}$ drift.
The gauge transformation
of electric field satisfying $\vctr{E}\cdot \vctr{B}=0$ enables this simple formulation.
In the previous data-driven simulations, the velocity fields inferred by DAVE4VM were introduced
in addition with
the inverted electric fields or the observational magnetic fields \citep{Hayashi2019ApJL,GuoYang2019ApJ}.
The DAVE4VM inversely solves the induction equation as well, whereas the vertical derivative
of the horizontal components of electric fields is assumed negligible.
In practice, the observed time evolution of the magnetic fields is not always 
reproduced completely even if the induction equation was integrated along with the DAVE4VM-inferred velocity fields.
Therefore, the previous data-driven simulations used the DAVE4VM-inferred velocity as the bottom boundary condition of the equation of motion, and the observational magnetic fields or the inverted electric fields
for the induction equation.
In the present simulation, the observational magnetic fields were reproduced
only by introducing the inverted velocity fields.
The bottom boundary condition for the equation of motion and the induction equation
is both physically and numerically consistent in our method.
We applied our method to the successive eruptive events.
Our simulation succeeded in reproducing the successive formation and eruption of flux ropes
as a response of temporal evolution of the observational magnetic fields.
The inversion technique of electric fields by \citet{Fisher2010ApJ} used in Step 1, described in Section \ref{sec:sim},
was widely used in the previous studies
\citep[][]{Pomoell2019SoPh}. The computation of the inverted velocity fields in Step 3 is straightforward.
Our method is simple yet feasible to reproduce magnetic activities in the solar atmosphere.

The energy release of flares in the solar active regions, the typical field strength of which is several thousand gauss, is in the range of $10^{28}-10^{32}~\mathrm{erg}$.
The field strength of the magnetic patches in our study was approximately one hundred gauss.
The estimation of the released energy of $10^{27}~\mathrm{erg}$ from our simulation result
is plausible for small eruptive events.

The successive flares and eruptions from active regions are often reported.
The largest flare in Solar Cycle 24, which marked X9.3 in GOES X-ray classification,
also had the preceding X2.2 flare. The flares were triggered by the continuous intrusion
of the opposite-polarity magnetic fluxes, 
according to the analyses by \citet{Bamba2020ApJ}.
In our case, although the spatial size and magnetic field strength were much smaller,
it is common that the continuous convergence of the opposite magnetic flux
and the subsequent partial deformation of the PIL were the triggers of the successive eruptions. 
It is worthy to note that the triggering mechanism of successive eruptions
was similar over a wide range of different spatial and temporal scales.
The previous theoretical studies \citep{Kusano2012ApJ,Kusano2020Sci} also support that the partial deformation of PILs can trigger eruption. \citet{Kusano2020Sci} discussed how a small area of reconnection can lead to MHD instability.  We speculate that the local converging motion can create small reconnetion-favor (opposite-polarity or reversed-shear type) regions along the PILs in the active regions.
In contrast, the origin of continuous converging motion is still unclear.
It is also unclear whether the converging motion is concentrated nearby PILs or ubiquitous in the photosphere.
The ultra high resolution observations by the Daniel K. Inouye Solar Telescope may reveal this issue.
A comprehensive understanding of the photospheric motion coupling with magnetic activity
in the convection zone is also required \citep[][]{Cheung2019NatAs,Hotta2019SciA,Toriumi2019ApJ}.

The onset times of the eruptions in our simulations were differed by $40~\mathrm{min}$ 
compared with the observational ones.
A possible reason is that the small-scale structures were smoothed out in our simulation.
As mentioned in Section \ref{sec:res}, the SSIMs of the magnetic fields after applying
the low-pass filter were already low.
The smaller structures may have to be included as much as possible to reproduce the accurate onset times.
The anomalous resistivity might also affect the results.
A parameter survey on the anomalous resistivity must be conducted in future work.

\acknowledgments
We are grateful to the anonymous referee for the constructive and thoughtful comments. This work was supported by MEXT/JSPS KAKENHI grant number JP15H05814, Project for Solar-Terrestrial Environmental Prediction (PSTEP), and JSPS KAKENHI grant number JP20K14519. This work was partially supported by MEXT as "Program for Promoting Researches on the Supercomputer Fugaku" (Toward a unified view of the universe: from large scale structure to planets, Elucidation of solar and planetary dynamics and evolution).
Numerical computations were conducted on a Cray XC50 supercomputer at the Center for Computational Astrophysics (CfCA)
of the National Astronomical Observatory of Japan. A part of this study was carried out using the computational resource of
the Center for Integrated Data Science, Institute for Space-Earth Environmental Research, Nagoya University.
HMI is an instrument on the SDO, a mission for NASA's Living with a Star program. We are grateful to the staff of Hida
Observatory for supporting the instrument development and daily observations.



\begin{thebibliography}{}
\expandafter\ifx\csname natexlab\endcsname\relax\def\natexlab#1{#1}\fi
\providecommand{\url}[1]{\href{#1}{#1}}
\providecommand{\dodoi}[1]{doi:~\href{http://doi.org/#1}{\nolinkurl{#1}}}
\providecommand{\doeprint}[1]{\href{http://ascl.net/#1}{\nolinkurl{http://ascl.net/#1}}}
\providecommand{\doarXiv}[1]{\href{https://arxiv.org/abs/#1}{\nolinkurl{https://arxiv.org/abs/#1}}}

\bibitem[{{Amari} {et~al.}(2014){Amari}, {Canou}, \& {Aly}}]{Amari2014Natur}
{Amari}, T., {Canou}, A., \& {Aly}, J.-J. 2014, \nat, 514, 465,
  \dodoi{10.1038/nature13815}

\bibitem[{{Bamba} {et~al.}(2020){Bamba}, {Inoue}, \& {Imada}}]{Bamba2020ApJ}
{Bamba}, Y., {Inoue}, S., \& {Imada}, S. 2020, \apj, 894, 29,
  \dodoi{10.3847/1538-4357/ab85ca}

\bibitem[{{Chae} {et~al.}(1998){Chae}, {Wang}, {Lee}, {Goode}, \&
  {Sch{\"u}hle}}]{1998ApJ...504L.123C}
{Chae}, J., {Wang}, H., {Lee}, C.-Y., {Goode}, P.~R., \& {Sch{\"u}hle}, U.
  1998, \apjl, 504, L123, \dodoi{10.1086/311583}

\bibitem[{{Cheung} \& {DeRosa}(2012)}]{CheungDerosa2012ApJ}
{Cheung}, M. C.~M., \& {DeRosa}, M.~L. 2012, \apj, 757, 147,
  \dodoi{10.1088/0004-637X/757/2/147}

\bibitem[{{Cheung} {et~al.}(2015){Cheung}, {De Pontieu}, {Tarbell}, {Fu},
  {Tian}, {Testa}, {Reeves}, {Mart{\'\i}nez-Sykora}, {Boerner}, {W{\"u}lser},
  {Lemen}, {Title}, {Hurlburt}, {Kleint}, {Kankelborg}, {Jaeggli}, {Golub},
  {McKillop}, {Saar}, {Carlsson}, \& {Hansteen}}]{CheungDerosa2015ApJ}
{Cheung}, M. C.~M., {De Pontieu}, B., {Tarbell}, T.~D., {et~al.} 2015, \apj,
  801, 83, \dodoi{10.1088/0004-637X/801/2/83}

\bibitem[{{Cheung} {et~al.}(2019){Cheung}, {Rempel}, {Chintzoglou}, {Chen},
  {Testa}, {Mart{\'\i}nez-Sykora}, {Sainz Dalda}, {DeRosa}, {Malanushenko},
  {Hansteen}, {De Pontieu}, {Carlsson}, {Gudiksen}, \&
  {McIntosh}}]{Cheung2019NatAs}
{Cheung}, M.~C.~M., {Rempel}, M., {Chintzoglou}, G., {et~al.} 2019, Nature
  Astronomy, 3, 160, \dodoi{10.1038/s41550-018-0629-3}

\bibitem[{{Fisher} {et~al.}(2010){Fisher}, {Welsch}, {Abbett}, \&
  {Bercik}}]{Fisher2010ApJ}
{Fisher}, G.~H., {Welsch}, B.~T., {Abbett}, W.~P., \& {Bercik}, D.~J. 2010,
  \apj, 715, 242, \dodoi{10.1088/0004-637X/715/1/242}

\bibitem[{{Guo} {et~al.}(2019){Guo}, {Xu}, {Ding}, {Chen}, {Xia}, \&
  {Keppens}}]{GuoYang2019ApJ}
{Guo}, Y., {Xu}, Y., {Ding}, M.~D., {et~al.} 2019, \apjl, 884, L1,
  \dodoi{10.3847/2041-8213/ab4514}

\bibitem[{{Hayashi} {et~al.}(2018){Hayashi}, {Feng}, {Xiong}, \&
  {Jiang}}]{Hayashi2018ApJ}
{Hayashi}, K., {Feng}, X., {Xiong}, M., \& {Jiang}, C. 2018, \apj, 856, 181,
  \dodoi{10.3847/1538-4357/aab787}

\bibitem[{{Hayashi} {et~al.}(2019){Hayashi}, {Feng}, {Xiong}, \&
  {Jiang}}]{Hayashi2019ApJL}
---. 2019, \apjl, 871, L28, \dodoi{10.3847/2041-8213/aaffcf}

\bibitem[{{He} {et~al.}(2020){He}, {Jiang}, {Zou}, {Duan}, {Feng}, {Zuo}, \&
  {Wang}}]{He2020ApJ}
{He}, W., {Jiang}, C., {Zou}, P., {et~al.} 2020, \apj, 892, 9,
  \dodoi{10.3847/1538-4357/ab75ab}

\bibitem[{{Hood} \& {Priest}(1979)}]{HoodPriest1979SoPh}
{Hood}, A.~W., \& {Priest}, E.~R. 1979, \solphys, 64, 303,
  \dodoi{10.1007/BF00151441}

\bibitem[{{Hotta} {et~al.}(2019){Hotta}, {Iijima}, \& {Kusano}}]{Hotta2019SciA}
{Hotta}, H., {Iijima}, H., \& {Kusano}, K. 2019, Science Advances, 5, 2307,
  \dodoi{10.1126/sciadv.aau2307}

\bibitem[{{Ichimoto} {et~al.}(2017){Ichimoto}, {Ishii}, {Otsuji}, {Kimura},
  {Nakatani}, {Kaneda}, {Nagata}, {UeNo}, {Hirose}, {Cabezas}, \&
  {Morita}}]{Ichimoto2017_SDDI_SoPh}
{Ichimoto}, K., {Ishii}, T.~T., {Otsuji}, K., {et~al.} 2017, \solphys, 292, 63,
  \dodoi{10.1007/s11207-017-1082-7}

\bibitem[{{Inoue}(2016)}]{Inoue2016PEPS}
{Inoue}, S. 2016, Progress in Earth and Planetary Science, 3, 19,
  \dodoi{10.1186/s40645-016-0084-7}

\bibitem[{{Ishiguro} \& {Kusano}(2017)}]{IshiguroKusano2017ApJ}
{Ishiguro}, N., \& {Kusano}, K. 2017, \apj, 843, 101,
  \dodoi{10.3847/1538-4357/aa799b}

\bibitem[{Jameson(2017)}]{Jameson2017origins}
Jameson, A. 2017, AIAA Journal, 1487

\bibitem[{{Jiang} {et~al.}(2016){Jiang}, {Wu}, {Yurchyshyn}, {Wang}, {Feng}, \&
  {Hu}}]{Jiang2016ApJ}
{Jiang}, C., {Wu}, S.~T., {Yurchyshyn}, V., {et~al.} 2016, \apj, 828, 62,
  \dodoi{10.3847/0004-637X/828/1/62}

\bibitem[{{Kliem} \& {T{\"o}r{\"o}k}(2006)}]{KliemTorok2006PhRvL}
{Kliem}, B., \& {T{\"o}r{\"o}k}, T. 2006, \prl, 96, 255002,
  \dodoi{10.1103/PhysRevLett.96.255002}

\bibitem[{{Kusano} {et~al.}(2012){Kusano}, {Bamba}, {Yamamoto}, {Iida},
  {Toriumi}, \& {Asai}}]{Kusano2012ApJ}
{Kusano}, K., {Bamba}, Y., {Yamamoto}, T.~T., {et~al.} 2012, \apj, 760, 31,
  \dodoi{10.1088/0004-637X/760/1/31}

\bibitem[{{Kusano} {et~al.}(2020){Kusano}, {Iju}, {Bamba}, \&
  {Inoue}}]{Kusano2020Sci}
{Kusano}, K., {Iju}, T., {Bamba}, Y., \& {Inoue}, S. 2020, Science, 369, 587,
  \dodoi{10.1126/science.aaz2511}

\bibitem[{{Kusano} {et~al.}(2002){Kusano}, {Maeshiro}, {Yokoyama}, \&
  {Sakurai}}]{Kusano2002ApJ}
{Kusano}, K., {Maeshiro}, T., {Yokoyama}, T., \& {Sakurai}, T. 2002, \apj, 577,
  501, \dodoi{10.1086/342171}

\bibitem[{{Maehara} {et~al.}(2012){Maehara}, {Shibayama}, {Notsu}, {Notsu},
  {Nagao}, {Kusaba}, {Honda}, {Nogami}, \& {Shibata}}]{Maehara2012Natur}
{Maehara}, H., {Shibayama}, T., {Notsu}, S., {et~al.} 2012, \nat, 485, 478,
  \dodoi{10.1038/nature11063}

\bibitem[{{Muhamad} {et~al.}(2017){Muhamad}, {Kusano}, {Inoue}, \&
  {Shiota}}]{Muhamad2017ApJ}
{Muhamad}, J., {Kusano}, K., {Inoue}, S., \& {Shiota}, D. 2017, \apj, 842, 86,
  \dodoi{10.3847/1538-4357/aa750e}

\bibitem[{{Namekata} {et~al.}(2020){Namekata}, {Maehara}, {Sasaki}, {Kawai},
  {Notsu}, {Kowalski}, {Allred}, {Iwakiri}, {Tsuboi}, {Murata}, {Niwano},
  {Shiraishi}, {Adachi}, {Iida}, {Oeda}, {Honda}, {Tozuka}, {Katoh}, {Onozato},
  {Okamoto}, {Isogai}, {Kimura}, {Kojiguchi}, {Wakamatsu}, {Tampo}, {Nogami},
  \& {Shibata}}]{Namekata2020PASJ}
{Namekata}, K., {Maehara}, H., {Sasaki}, R., {et~al.} 2020, \pasj, 72, 68,
  \dodoi{10.1093/pasj/psaa051}

\bibitem[{{Notsu} {et~al.}(2019){Notsu}, {Maehara}, {Honda}, {Hawley},
  {Davenport}, {Namekata}, {Notsu}, {Ikuta}, {Nogami}, \&
  {Shibata}}]{Notsu2019ApJ}
{Notsu}, Y., {Maehara}, H., {Honda}, S., {et~al.} 2019, \apj, 876, 58,
  \dodoi{10.3847/1538-4357/ab14e6}

\bibitem[{{Osten} {et~al.}(2005){Osten}, {Hawley}, {Allred}, {Johns-Krull}, \&
  {Roark}}]{Osten2005ApJ}
{Osten}, R.~A., {Hawley}, S.~L., {Allred}, J.~C., {Johns-Krull}, C.~M., \&
  {Roark}, C. 2005, \apj, 621, 398, \dodoi{10.1086/427275}

\bibitem[{{Pandey} \& {Singh}(2008)}]{PandeySingh2008MNRAS}
{Pandey}, J.~C., \& {Singh}, K.~P. 2008, \mnras, 387, 1627,
  \dodoi{10.1111/j.1365-2966.2008.13342.x}

\bibitem[{{Pesnell} {et~al.}(2012){Pesnell}, {Thompson}, \&
  {Chamberlin}}]{Pesnell2012_SDO_SoPh}
{Pesnell}, W.~D., {Thompson}, B.~J., \& {Chamberlin}, P.~C. 2012, \solphys,
  275, 3, \dodoi{10.1007/s11207-011-9841-3}

\bibitem[{{Pomoell} {et~al.}(2019){Pomoell}, {Lumme}, \&
  {Kilpua}}]{Pomoell2019SoPh}
{Pomoell}, J., {Lumme}, E., \& {Kilpua}, E. 2019, \solphys, 294, 41,
  \dodoi{10.1007/s11207-019-1430-x}

\bibitem[{{Priest}(2014)}]{Priest2014}
{Priest}, E. 2014, {Magnetohydrodynamics of the Sun} (Cambridge University
  Press), \dodoi{10.1017/CBO9781139020732}

\bibitem[{{Rempel}(2014)}]{Rempel2014ApJ}
{Rempel}, M. 2014, \apj, 789, 132, \dodoi{10.1088/0004-637X/789/2/132}

\bibitem[{{Schou} {et~al.}(2012){Schou}, {Scherrer}, {Bush}, {Wachter},
  {Couvidat}, {Rabello-Soares}, {Bogart}, {Hoeksema}, {Liu}, {Duvall}, {Akin},
  {Allard}, {Miles}, {Rairden}, {Shine}, {Tarbell}, {Title}, {Wolfson},
  {Elmore}, {Norton}, \& {Tomczyk}}]{Schou2012_HMI_SoPh}
{Schou}, J., {Scherrer}, P.~H., {Bush}, R.~I., {et~al.} 2012, \solphys, 275,
  229, \dodoi{10.1007/s11207-011-9842-2}

\bibitem[{{Schuck}(2006)}]{Schuck2006ApJ}
{Schuck}, P.~W. 2006, \apj, 646, 1358, \dodoi{10.1086/505015}

\bibitem[{{Toriumi} {et~al.}(2020){Toriumi}, {Takasao}, {Cheung}, {Jiang},
  {Guo}, {Hayashi}, \& {Inoue}}]{Toriumi2020ApJ}
{Toriumi}, S., {Takasao}, S., {Cheung}, M. C.~M., {et~al.} 2020, \apj, 890,
  103, \dodoi{10.3847/1538-4357/ab6b1f}

\bibitem[{{Toriumi} \& {Hotta}(2019)}]{Toriumi2019ApJ}
{Toriumi}, S., \& {Hotta}, H. 2019, \apjl, 886, L21,
  \dodoi{10.3847/2041-8213/ab55e7}

\bibitem[{{UeNo} {et~al.}(2004){UeNo}, {Nagata}, {Kitai}, {Kurokawa}, \&
  {Ichimoto}}]{Ueno2004_SMART_SPIE}
{UeNo}, S., {Nagata}, S.-i., {Kitai}, R., {Kurokawa}, H., \& {Ichimoto}, K.
  2004, in Society of Photo-Optical Instrumentation Engineers (SPIE) Conference
  Series, Vol. 5492, Ground-based Instrumentation for Astronomy, ed. A.~F.~M.
  {Moorwood} \& M.~{Iye}, 958--969, \dodoi{10.1117/12.550304}

\bibitem[{{Wang} {et~al.}(1998){Wang}, {Johannesson}, {Stage}, {Lee}, \&
  {Zirin}}]{1998SoPh..178...55W}
{Wang}, H., {Johannesson}, A., {Stage}, M., {Lee}, C., \& {Zirin}, H. 1998,
  \solphys, 178, 55, \dodoi{10.1023/A:1004974927114}

\bibitem[{Wang {et~al.}(2004)Wang, Bovik, Sheikh, \&
  Simoncelli}]{Wang2004image}
Wang, Z., Bovik, A.~C., Sheikh, H.~R., \& Simoncelli, E.~P. 2004, IEEE
  transactions on image processing, 13, 600

\end{thebibliography}



\end{document}